# VWATTACKER: A Systematic Security Testing Framework for Voice over WiFi User Equipments

Imtiaz Karim*
The University of Texas at Dallas
imtiaz.karim@utdallas.edu

Hyunwoo Lee*
Korea Institute of Energy Technology
hwlee@kentech.ac.kr

Hassan Asghar
Korea Institute of Energy Technology
hassanasghar@kentech.ac.kr

Kazi Samin Mubasshir
Purdue University
kmubassh@purdue.edu

Seulgi Han
Korea Institute of Energy Technology
ab9938@kentech.ac.kr

Mashroor Hasan Bhuiyan
The University of Texas at Dallas
mashroorhasan.bhuiyan@utdallas.edu

Elisa Bertino
Purdue University
bertino@purdue.edu

## ABSTRACT

We present VWATTACKER, the *first systematic testing* framework for analyzing the security of Voice over WiFi (VoWiFi) User Equipment (UE) implementations. VWATTACKER includes a complete VoWiFi network testbed that communicates with Commercial-Off-The-Shelf (COTS) UEs based on a simple interface to test the behavior of diverse VoWiFi UE implementations; uses property-guided adversarial testing to uncover security issues in different UEs systematically. To reduce manual effort in extracting and testing properties, we introduce an LLM-based, semi-automatic, and scalable approach for property extraction and testcase (TC) generation. These TCs are systematically mutated by two domain-specific transformations. Furthermore, we introduce two deterministic oracles to detect property violations automatically. Coupled with these techniques, VWATTACKER extracts 63 properties from 11 specifications, evaluates 1,116 testcases, and detects 13 issues in 21 UEs. The issues range from enforcing a DH shared secret to 0 to supporting weak algorithms. These issues result in attacks that expose the victim UE's identity or establish weak channels, thus severely hampering the security of cellular networks. We responsibly disclose the findings to all the related vendors. At the time of writing, one of the vulnerabilities has been acknowledged by MediaTek with high severity.

## 1 INTRODUCTION

Voice over WiFi (VoWiFi) is a mobile call service that aims to extend network coverage by utilizing WiFi in areas where a cellular network is unavailable or the cellular signal is weak. Poor signal is not a new issue because typically areas within buildings or underground areas are not covered by conventional mobile networks [12, 13]. Furthermore, around 80% of all calls are usually made indoors, according to one survey [14]. Due to these limitations of the current cellular network infrastructure, VoWiFi usage is increasing with the large number of WiFi networks deployed in homes/offices and the growth of public WiFi [17]. The VoWiFi market is estimated to reach USD 21.99 billion in 2030 from USD 9.59 billion expected in 2025 with an annual growth rate of more than 18% [2]. Such widespread use of VoWiFi requires robust security guarantees.

**Prior research and scope.** Although previous works have analyzed the security of the VoWiFi protocol specifications [10, 40, 44, 54, 60], VoLTE specifications and implementations [38, 43], and cellular network implementations in general [22, 28, 39, 50, 57], there is no systematic framework for analyzing VoWiFi implementations with a complete open-source VoWiFi commercial UE testbed (see Table 1). Furthermore, the main body of work [10, 44, 60] on VoWiFi looks at scattered parts of the protocol, and the approaches are ad hoc and manual. Lee *et al.* [40] verify the complete protocol based on formal verification, but their analysis focuses only on the specifications. Concerning the security analysis of VoWiFi implementations, our only reference is the conformance test suites [6] provided by 3GPP. The conformance test suites aim to ensure that VoWiFi implementations meet the minimum requirements. However, these testcases (TCs) have limitations. They (i) focus on functional requirements rather than security, and (ii) do not evaluate the behaviors of VoWiFi implementations in adversarial scenarios. Although there has been a broad body of work on cellular implementation testing [22, 39, 50], those frameworks are built on top of different open-source cellular network implementations and are not general enough to include VoWiFi testing.

Motivated by this significant gap in the testing of VoWiFi implementations, we developed the first *systematic* security testing framework–VWATTACKER to evaluate the security of VoWiFi UE implementations. At a high level, VWATTACKER is an adversarial property-guided testing framework. However, we include significant innovations to reduce manual effort and substantially increase the scope of testing.

**Challenges.** The first challenge for the adversarial testing on COTS UEs is that there is no open-source VoWiFi testbed that controls and synchronizes the many entities involved in VoWiFi communications (e.g., UEs, the ePDG, and the IMS). The second challenge is related to the scalability of property-guided adversarial testing. In this broader challenge, we face two subchallenges. First, such a testing technique requires significant manual effort to extract properties from the specifications, and this limits comprehensive security testing [39, 50]. Second, it is difficult to detect when a property violation has occurred without any manual intervention. As the specifications are written in natural language and contain multiple underspecifications [40], it is not always possible to infer the correct behavior.

---
*The authors contributed equally to this research.

| Paper | Systematic Frame-work | Implementation Analysis | Complete VoWiFi UE Testbed | Scalable Property Extrac-tion |
|---|---|---|---|---|
| Wi-Not-Calling [10] | | ✓ | | |
| Xie et al. [60] | | ✓ | | |
| Lu et al. [44] | | ✓ | | |
| Shi et al. [54] | | ✓ | | |
| Gegenhuber et al. [23] | | ✓ | | |
| VWANALYZE [40] | ✓ | | | |
| Li et al. [43] | ✓ | | | |
| Kim et al. [38] | ✓ | | | |
| DIKEUE [28] | ✓ | ✓ | | |
| 5GBaseChecker [57] | ✓ | ✓ | | |
| LTEFuzz [39] | ✓ | ✓ | | |
| DoLTEst [50] | ✓ | ✓ | | |
| VWATTACKER | ✓ | ✓ | ✓ | ✓ |

**Table 1: Comparison of VWATTACKER with other analyses on cellular networks, VoWiFi, and VoLTE**

**Approach.** To address the first challenge and bridge the gap for testing VoWiFi UE implementations, VWATTACKER includes a complete VoWiFi network testbed that we built based on open-source software (*StrongSwan* [55] for ePDG, *Kamailio* [33] for IMS, and *FHoSS* [26] for Home Subscriber Server (HSS)). The testbed is currently developed for 4G LTE because, with regards to VoWiFi, 4G LTE is still the most dominant technology [25]. To control VoWiFi entities, we design a *command-report* protocol based on a central-peripheral paradigm, where a central controller manages communication between all entities. We also implement a simple JSON-based interface that allows testers to describe any sequence of messages containing specific attribute values, enabling us to conduct efficient adversarial testing.

In a recent independent and concurrent project, Osmocom has developed an open-source *osmo-ePDG* for VoWiFi UE connectivity [49]. Such an effort not only attests to the importance of the problem but also motivates its urgency. Though promising, osmo-ePDG has several key differences from our testbed. First, osmo-ePDG is a testbed just for VoWiFi UE connectivity and not for security testing. In contrast, our testbed is designed for security testing, featuring a JSON-based testing format for describing adversarial and out-of-order messages to be sent to UEs, as well as a separate control architecture for facilitating easier testing. Second, osmo-ePDG is a module extending the osmocom mobile network, while ours is dedicated to the VoWiFi security analysis with minimal additional functions.

To address the second challenge and enhance property-guided testing, we design TESTGEN that adopts a scalable and semi-automatic Large Language Model (LLM)-based property extraction approach to extract diverse properties. We leverage the *in-context learning* capability of the LLMs to learn from the VoWiFi specifications and generate properties using Retrieval Augmented Generation (RAG) [42]. With just a few example properties and the VoWiFi specifications as context, an LLM can accurately generate additional properties and encode them as Primary Test Cases (PTCs), eliminating costly manual property extraction. These PTCs are then mutated using the two transformations, creating Adversarial Test Cases (ATCs) and sent to the testbed. As messages that contain attributes are exchanged in the VoWiFi protocol, we design two types of transformations: (i) *message-level* transformation inserts, replaces, drops, or replays a target *message*; and (ii) *attribute-level* transformation inserts, updates, or drops an *attribute* in a target message. To automatically detect property violations, we design two different oracles-*function* and *liveness* oracles, where the former detects invalid functions of UEs, and the latter identifies deadlock states of UEs.

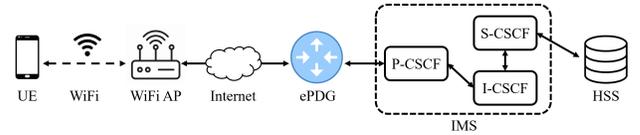

**Figure 1: VoWiFi architecture.**

**Results.** With VWATTACKER, we evaluate 21 UEs from 14 different UE vendors and 5 baseband vendors with Android versions ranging from 7 to 14. For the property testing, TESTGEN extracts 63 PTCs from 11 VoWiFi-related specifications. After applying two types of transformations, we get 1,116 ATCs. We report 13 unique issues. Notable among them are issues where implementations: (i) support and accept weak/deprecated algorithms (e.g., DES or MD5); and (ii) accept packets without a nonce or a DH key that can be used to bypass the DH key security to expose the devices' IMSI; The impact of these attacks ranges from privacy leaks to establishing weak channels with weak encryption and integrity algorithms, severely hampering the security of the implementation.

**Open-source.** We have open-sourced VWATTACKER, including the complete VoWiFi network testbed, the LLM-based property extraction technique, and the oracles to foster research in this area and help vendors test their UEs [*].

**Responsible Disclosure.** As our findings can result in the real-world attacks, we have responsibly disclosed the findings of our work to all the related vendors (i.e., baseband vendors and UE vendors) and are actively cooperating with them for mitigation. At the point of the writeup one issue has been acknowledged by MediaTek with high-severity.

**Contributions.** We summarize our contributions as follows:

- We propose VWATTACKER, the framework for analyzing the security of VoWiFi implementations. VWATTACKER includes a *complete* VoWiFi network testbed for COTS UEs. To the best of our knowledge, VWATTACKER is the *first* systematic security framework to test VoWiFi UE implementations.
- To reduce manual labor and increase the scalability of property-guided testing, we design TESTGEN, an LLM-based semi-automatic property extraction technique based on in-context learning and RAG. Using this approach, VWATTACKER extracts 63 properties from the 11 specifications, which results in 1,116 ATCs after applying two types of transformations.
- We test a total of 21 COTS UEs from 14 vendors. Our deterministic oracles uncover 13 issues. Based on vulnerabilities, we reveal 3 new attacks.

## 2 VOICE OVER WIFI

This section provides an overview of the VoWiFi protocol. We first present the VoWiFi architecture (see Figure 1) followed by the important subprotocols during the procedure of the UE registration (see Figure 2). They are the Internet Key Exchange version 2 (IKEv2) [34], the Extensible Authentication Protocol Method for 3rd Generation Authentication and the Key Agreement (EAP-AKA) [9], and the Session Initiation Protocol (SIP) [53].

---

[*]https://github.com/hw5773/vowifi-ue-testing-framework

**VoWiFi Architecture.** VoWiFi is a Voice over IP (VoIP) service run by a mobile network. To use the service, the *User Equipment* (UE) sets up a Virtual Private Network (VPN) with the mobile network after it attaches to a WiFi access point (AP). The VPN peer that communicates with the UE on the mobile network side is called an *evolved Packet Data Gateway* (ePDG). It is a connection point between the Internet and the mobile network. Once the VPN is established, the UE starts communicating with the *IP Multimedia Subsystem* (IMS), the main components of which are the *Proxy Call Session Control Function* (P-CSCF), the *Interrogating Call Session Control Function* (I-CSCF), and the *Service Call Session Control Function* (S-CSCF). The P-CSCF is an endpoint that directly communicates with the UE by exchanging the SIP messages. It forwards a request from the UE to the I-CSCF that is responsible for selecting the S-CSCF to be assigned for the session. The S-CSCF is mainly responsible for registration. In the VoWiFi architecture (i.e., UE, ePDG, and IMS) as *VoWiFi entities* (or *entities*). These entities execute the IKE protocol, the EAP-AKA protocol, and the SIP protocol, respectively.

**IKEv2 [34].** It is the subprotocol used to establish the VPN between the UE and an ePDG. The VPN is set up with a common security association, including cryptographic algorithms and keys. IKEv2 proceeds with several *exchanges*; each such exchange consists of a *request* message and a *response* message. In other words, a request message sent by one entity is always followed by a response message from the other entity. In the VoWiFi scenario, the UE always initiates IKEv2 by sending the `IKE_SA_INIT` request message to the ePDG, which responds with the `IKE_SA_INIT` response message. The `IKE_SA_INIT` exchange is responsible for establishing the IKE security association (SA) that contains cryptographic keys and algorithms. Then, the protocol is followed by several `IKE_AUTH` exchanges that are responsible for authenticating each other through EAP-AKA and establishing a child SA. Note that these exchanges are secured with the `IKE_SA`.

**EAP-AKA [9].** It is the subprotocol used for mutual authentication between the UE and the ePDG, utilizing a challenge-response mechanism and symmetric cryptography. The UE and the ePDG exchange AKA messages encapsulated with the `IKE_AUTH` exchanges. In the VoWiFi protocol, EAP-AKA is initiated after the ePDG receives the first `IKE_AUTH` request message from the UE, which includes its identifier (i.e., International Mobile Subscriber Identity, IMSI) and the intended peer's identifier (i.e., ims). The ePDG fetches the key materials through the UE's IMSI and responds with the `EAP-Request/AKA-Challenge` message that contains a random number (i.e., `AT_RAND`), an authentication token (i.e., `AT_AUTN`), and a message authentication code (i.e., `AT_MAC`). The UE verifies both `AT_AUTN` and `AT_MAC` and sends `AKA-Client-Error` if any of them is invalid. Otherwise, the UE encrypts the random number with its unique key and sends the encrypted message (i.e., `AT_RES`) with a message authentication code (i.e., `AT_MAC`) in the `EAP-Response/AKA-Challenge` message. If the ePDG successfully validates `AT_RES`, which means that the UE is authenticated, the ePDG sends the `AKA-Success` message to the UE, and the IKE channel is finally established between the UE and the ePDG.

**SIP [53].** It is the subprotocol used to register a UE with the IMS or to make and receive a VoWiFi call. As we focus on the SIP

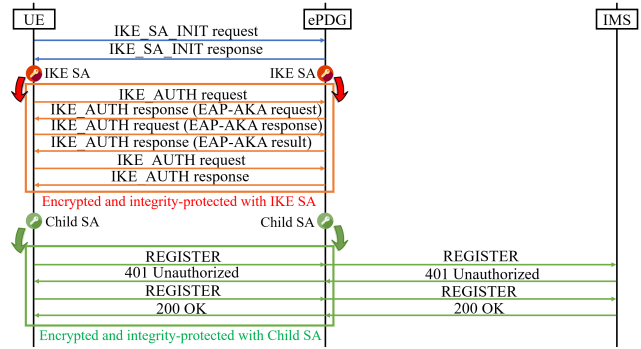

**Figure 2: Flow of the VoWiFi protocol**

registration in this paper, we only describe the registration flow below, which is based on the challenge-response protocol. Once the IKE channel is established, the UE sends the `SIP REGISTER` message to the IMS server through the ePDG. IMS responds with the `401 Unauthorized` message that includes a nonce value. The UE generates the authentication token with the nonce and responds with the second `SIP REGISTER` message which contains the token. Finally, the UE is authenticated and registered if the token is valid, and IMS sends a confirmation message (i.e., `200 OK`). Note that the SIP messages are encrypted and integrity-protected with the child SA established during the IKE protocol.

## 3 OVERVIEW OF VWATTACKER

This section provides an overview of the VWAttacker (see Figure 3) with its threat model, challenges, and requirements.

### 3.1 Threat Model

In adversarial testing, we assume a Dolev-Yao attacker [20]. The attacker can eavesdrop, modify, or drop any message and inject new messages. However, the attacker is computationally bounded; that is, the attacker cannot break cryptographic assumptions. Therefore, the attacker cannot decrypt an encrypted message unless they possess the decryption key. This is a realistic threat model to uncover protocol-related issues and is heavily used by different protocol testing frameworks [11, 16, 18, 27, 50, 59].

In the context of VoWiFi, we assume that the attacker resides on the communication path between the VoWiFi UE and the VoWiFi core network. This includes malicious WiFi access points, compromised routers, or untrusted intermediaries that can intercept and manipulate network traffic. For instance, the attacker can impersonate a legitimate WiFi network (e.g., Evil Twin AP) to intercept VoWiFi signaling and media traffic [7, 40, 41, 47, 51]. However, the attacker does not compromise the UE itself and cannot access protected memory or internal protocol state of the VoWiFi implementation.

### 3.2 Challenges and Requirements

The critical challenges in the design of VWAttacker are:

**(C1) VoWiFi network testbed.** As commercial VoWiFi UEs are entirely black-box, designing a security testing framework requires a complete UE testbed to interface with them. Unfortunately, no open-source VoWiFi UE testbed is available for us to use. This is a fundamental challenge that prevents VoWiFi implementation testing. Furthermore, the testbed should be fully customizable to implement a systematic framework for adversarial testing and should allow UEs to execute the complete VoWiFi procedures.

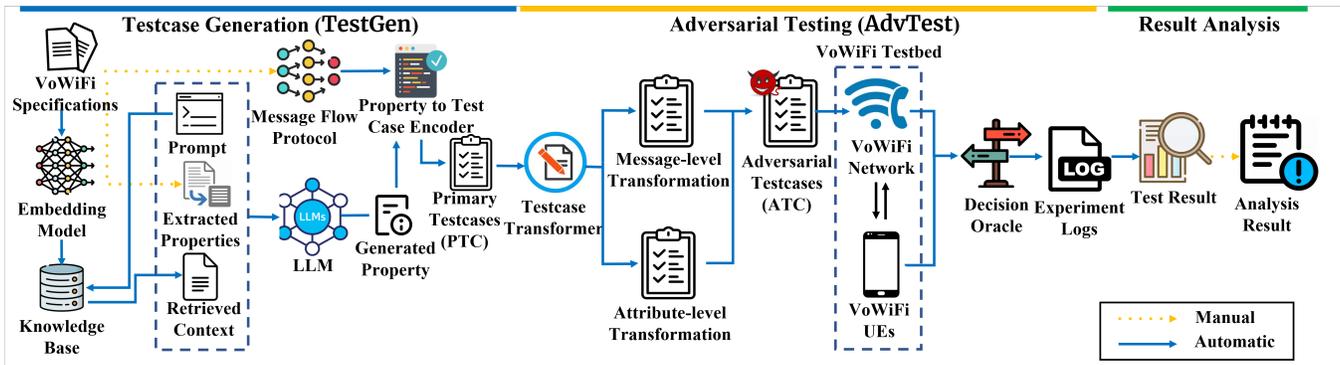

**Figure 3: Overview of VWAttacker.**

**(C2) Scalability of property-guided testing.** To comprehensively analyze the VoWiFi implementations and improve the scalability of property-guided testing, we need to tackle two sub-challenges:

- **(C2.1) Scalable and accurate property extraction:** Previous works on property-based adversarial testing rely on manual property extraction [39, 50], which is time-consuming, error-prone, and limited in scope. Although recent work explores LLM-assisted fuzzing [45] for other domains, these approaches typically lack integration with wireless protocol semantics or are optimized for application-layer protocols. Moreover, they do not address the ambiguities or underspecifications common in VoWiFi standards, which can lead to hallucinations. Hence, a specialized approach is needed to extract low-level protocol properties from fragmented specifications (3GPP, RFCs, conformance suites) while preserving contextual correctness and completeness.

- **(C2.2) Property-violation detection:** Another crucial part of property-guided testing is detecting when the property has been violated. As the VoWiFi protocol does not include a formal canonical model, it is impossible to directly compare the behavior with a canonical model to detect deviations and property violations. Therefore, we must design side-channel oracles that automatically detect property violations.

### 3.3 Addressing the Challenges

To address (C1), we build a VoWiFi network testbed that provides commercial UEs with VoWiFi service by integrating existing open-source subprotocol implementations, such as *StrongSwan* and *Kamailio*, together with a custom ISIM card, *sysmoISIM-SJA2*. In deploying a completely VoWiFi-supporting testbed, we aim to support diverse devices without requiring any modification on them (e.g., no rooting) and extra hardware. Due to these requirements, customizing the VoWiFi network needs considerable engineering effort. We detail how we develop the testbed in Section 4.

To address (C2.1), we design an LLM-based property extraction and testcase generation framework, called TestGen, that extracts diverse properties from the VoWiFi specifications and generates test-cases in a semi-automated fashion. LLMs can learn from in-context information and utilize this information to perform specific downstream tasks without requiring retraining or modifying the model weights. With VoWiFi specifications as context and some manually extracted properties as examples, we use an LLM to extract properties from VoWiFi specifications. Although modern LLMs are likely pre-trained on many publicly available protocol specifications, relying solely on their internal memory can lead to hallucinations or overlook domain-specific details scattered across loosely connected sections of the specifications. To tackle this challenge, we utilize Retrieval Augmented Generation (RAG). By explicitly grounding the model with relevant context at query time, RAG ensures that the LLM does not have to rely on memorization from massive pretraining corpora, but instead focuses on interpreting the specification fragments most relevant to the prompt. This makes the extraction process both more targeted and transparent, especially in a domain like VoWiFi where specifications are complex, modular, and often underspecified [40]. After extracting the properties, an encoder converts these extracted properties into the primary testcases (PTCs) using the message flows of related protocols. Two transformation techniques are then applied to these PTCs, resulting in adversarial testcases (ATCs), which are used to test the security of UEs in the customized VoWiFi testbed.

To address (C2.2) and automatically detect property violations, we design two types of decision oracles that retrieve logs generated from each test and raise alerts whenever they find issues. These are used to detect semantic bugs or availability issues.

### 3.4 High-Level Overview of VWAttacker

At a high level, VWAttacker is divided into three modules (see Figure 3): (i) testcase generation (TestGen); (2) adversarial testing (AdvTest); and (3) result analysis.

TestGen generates testcases in two parts:

- **(1) Property extraction using LLMs:** Initially, some properties are manually extracted and provided as examples to guide the LLM in understanding the type of properties to look for in the specifications. Using the provided context and examples, the LLM processes the VoWiFi specifications to identify and extract a comprehensive set of properties.

- **(2) Property-to-testcase encoding:** The properties extracted by the LLM are fed into the property-to-test-case encoder, along with the message flows of the VoWiFi protocol. The encoder translates the properties into PTCs, which are then used to automatically test the VoWiFi system in a dedicated testbed, ensuring comprehensive coverage and efficient validation.

The second step, ADVTEST, has three main parts:

- **(1) Testcase transformer:** We mutate the PTCs using the two transformations, creating ATCs. As messages containing attributes are exchanged in the main subprotocols of VoWiFi, namely IKEv2 and SIP, we design two types of transformations with mutation targets at different levels. The *message-level* transformation inserts, replaces, drops, or replays a target *message*, while the *attribute-level* transformation inserts, updates, or drops an *attribute* in a target message. After applying these transformations to PTCs, we get ATCs.
- **(2) VoWiFi UE testing:** ATCs are sent to the controller of the VoWiFi testbed. The controller processes them, directs VoWiFi entities to send messages according to ATCs, receives the responses and outputs the logs.
- **(3) Decision oracles:** The decision oracles retrieve logs and report "positives" if they find UEs' misbehavior due to adversarial testing or their deadlock states.

In the final step, result analysis, we run scripts to summarize the experiment logs into the test results. Then, we manually inspect the test results and investigate the root causes of UEs' misbehavior or deadlock states.

# 4 VOWIFI NETWORK TESTBED
In this section, we describe the VoWiFi network testbed.

## 4.1 Requirements
To support a comprehensive black-box property-guided adversarial testing framework for VoWiFi, the underlying testbed should satisfy the following requirements: **(R1) Device-diversity:** The testbed should be able to work with diverse UEs. Satisfying this requirement is challenging, as VoWiFi implementations show quite different behavior. For instance, some UEs always send the IKE_-delete message before turning off the WiFi interface, while others do not send the message. **(R2) Plug-and-play:** The testbed should not require any change in UEs and be plug-and-play. For instance, if the testbed were to require rooting the UEs, it would entail manual effort on the part of testers, and the usage of the testbed would be minimal. **(R3) Malleable testing system:** Once we have the VoWiFi network testbed, we can send any test input to VoWiFi UEs. To evaluate the security of VoWiFi UEs in a wide range of scenarios, the testbed should support testing any possible scenario provided by the tester. This, in turn, requires generating a stateless system from a highly stateful protocol. To make the testbed useful, all VoWiFi entities should be directly controlled, and log messages from these entities should be collected in a unified format.

## 4.2 High-level Design
To design the testbed, we integrate existing open-source subprotocol implementations, such as *StrongSwan* and *Kamailio*, together with a custom ISIM card, *sysmoISIM-SJA2*, and make 21 UEs from 14 different vendors runnable on the testbed satisfying **R1**. Our testbed operates in a plug-and-play fashion, requiring no modifications to the devices or additional hardware, following **R2**. We provide a simple, JSON-based interface for testers to describe how they want to evaluate a target UE. We define a testcase (TC) as a series of commands by which the tester can make VoWiFi entities send specific messages (e.g., making ePDG send the second IKE_AUTH response message with the value of the EAP failure in EAP-AKA) according to the test UE's messages. For creating the malleable testbed following **R3**, we design it based on a control architecture following the *central-peripheral* paradigm, where a controller directs VoWiFi entities (i.e., UE, ePDG, and IMS) through a UE agent, an ePDG agent, and an IMS agent, respectively. To achieve this, we move all the logic away from the peripheral agents to the CONTROLLER to create a stateless system.

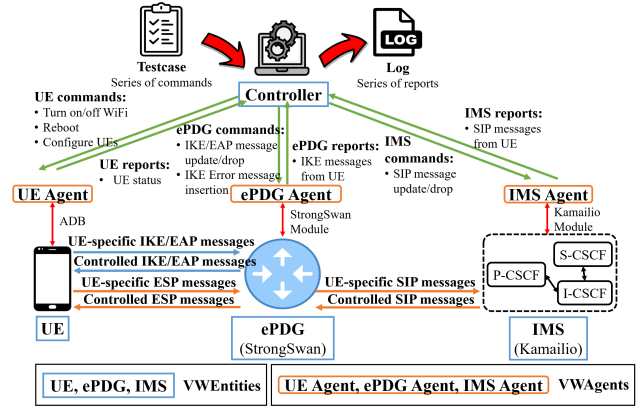

**Figure 4: The control architecture of the Testbed**

## 4.3 Control architecture
Our control architecture (see Figure 4) consists of a CONTROLLER that processes a TC and AGENTS (UEAGENT, EPDGAGENT, and IMSAGENT) that communicate with the CONTROLLER to control VoWiFi entities (i.e., UE, ePDG, and IMS). Based on the CONTROLLER and AGENTS, VWATTACKER takes a TC described by a series of commands as input and outputs a log, which is a series of reports from AGENTS. To direct VoWiFi entities according to a TC, we use a *command-report* protocol between the CONTROLLER and the AGENTS. In detail, the CONTROLLER sends a command to an agent to control the corresponding VoWiFi entity. Then, the agent reports the result of the command (e.g., UE's status or responding messages from UE) for which the agent is responsible to the CONTROLLER.

**Testcase.** A *testcase* (TC), shown in Figure 5, is an interface provided for a tester to describe the scenario of their interest encoded in a simple JSON format. In detail, a TC is a series of commands that describe *who* is responsible for sending *what* message, with *how* one inserts/modifies/drops a message or its attributes. Once the CONTROLLER gets a TC, it directs the corresponding agents to execute a sequence of commands described in the TC. The commands sent to all agents and the reports they submit are saved for further analysis. For a TC, we define 10 keywords as JSON keys in describing a TC. The keywords are used to indicate a receiver of a command (receiver), the name of a message to be altered and sent (name), an operation to be applied to the message (op), and others. With the

keywords, a tester, for instance, can describe a command to let the ePDG send an IKE_SA_INIT message after modifying the encryption algorithm to DES.

**Controller.** The Controller processes a TC while directing the VoWiFi entities to do their tasks through the Agents. We design and implement VWAttacker to take multiple TCs at once and process all the TCs one by one without human intervention. To process multiple TCs, it is required that all the states of the VoWiFi entities should be initialized before processing the next TC. To this end, VWAttacker refreshes all the VoWiFi entities before it runs the next TC. While processing the TCs, the Controller communicates with Agents to direct the corresponding VoWiFi entities to send messages or report results according to the TCs.

**UE agent.** Once a UE is attached to VWAttacker (i.e., by connecting the UE with our framework via the USB cable), the UEAgent checks the model name of the UE and configures the VoWiFi setup of the UE. This configuration is necessary because some UEs require toggling the WiFi call switch in the settings app to enable the VoWiFi service. Then, the UEAgent becomes ready to process commands from the Controller. The commands that UEAgent can process are: ❶ turning on/off the WiFi interface and ❷ rebooting the UE. The only report that UEAgent sends to the Controller is the result of the reboot command. This report is required as the Controller needs to know when a UE is ready to send a message so that the Controller can send a message to the agent.

**ePDG agent.** The ePDGAgent can substitute a message, reply to a message, insert an attribute-value pair, update the value of an attribute, and drop an attribute at the appropriate points of the original *StrongSwan* source code. The ePDGAgent can follow the following commands: ❶ sending an IKE message; ❷ substituting a message; ❸ replaying an IKE message; ❹ inserting an attribute-value pair; ❺ updating the value of an attribute; ❻ dropping a specific attribute. The ePDGAgent reports IKE messages, including headers and payloads, that the ePDG sends to and receives from the attached UE. These reported messages are logged into the output log file of VWAttacker.

**IMS agent.** Like the ePDGAgent, IMSAgent modifies, replays, updates, or drops a message according to the commands in TCs. The IMSAgent accepts the following commands: ❶ sending a SIP message; ❷ modifying a value; ❸ deleting a specific attribute-value pair.

## 5 DETAILS OF VWATTACKER

This section describes how we design and implement VWAttacker leveraging the VoWiFi testbed. Since the result analysis module is trivial, in the following subsections, we present the details of Test-Gen and AdvTest–which are the novel elements of VWAttacker..

### 5.1 TestGen: Testcase Generation

TestGen consists of two components: (1) the LLM-based property extractor and (2) the property-to-testcase encoder. The former contributes to extracting properties from specifications, while the latter encodes the properties into PTCs.

**LLM-based property extractor.** To automatically extract security-relevant properties from loosely structured VoWiFi specifications, we build a Retrieval-Augmented Generation (RAG) framework tailored for this domain. The framework begins by embedding 11 specification documents, including 3GPP technical reports, IETF RFCs and conformance test specifications. These embedded chunks are indexed for efficient semantic retrieval. We then use a retrieval module to fetch specification fragments relevant to a given query. Along with a small set of manually curated example properties (created in approximately 1 hour by a domain expert), these retrieved documents are provided as context to a prompted large language model (LLM). The model then generates candidate properties grounded in the retrieved content. Although LLMs are likely pretrained on publicly available specifications, relying on their internal knowledge can lead to inconsistent outputs, especially for rarely cited clauses or ambiguous conditions. Using RAG avoids these pitfalls by explicitly providing relevant specification fragments at inference time, making the extraction more precise, consistent, and explainable. Compared to fine-tuning or naive prompting, RAG offers a flexible and modular solution. It improves performance (see Section 7) without requiring retraining or labeled datasets, and enables us to trace each generated property back to specific specification fragments.

**Property-to-testcase encoder.** The encoder takes the generated properties, the RAG-based model used for property extraction and the protocol message flow as input and converts the properties into PTCs. To achieve this, the encoder first identifies the message name from the property and matches it to the protocol's message flow to determine the state required to execute the property. Then, following this state, the encoder creates a series of commands by appending messages with related VoWiFi entities one after another, based on the protocol's message flow. For each message, the relevant fields and values (if not specified in the property) are generated by querying the RAG-based LLM model. A high-level overview of the used protocol flow diagram is shown in Figure 6. Note that, not all the properties include an explicit message name and refers to message fields; in those cases, the message names are manually provided to the encoder.

### 5.2 AdvTest: Adversarial Testing

AdvTest begins with the testcase transformer that converts PTCs into ATCs. Then, ATCs are fed into the VoWiFi testbed for the adversarial testing. Finally, the misbehavior of UEs is automatically detected with the help of the decision oracles.

**Testcase transformer.** For generating the ATCs, we develop transformations at two different levels: message-level and attribute-level. A *message-level transformation* changes the final command of the PTC to generate an adversarial message. There are two types of message-level transformations: ❶ **Message substitution:** the final command is set to send an error message (e.g., an invalid SPI) to see how a UE reacts to unexpected error messages. ❷ **Message replay:** We manipulate a final command to send one of the previous messages again to check how a UE behaves against the replay attack.

Next, an *attribute-level transformation* is to modify or remove one specific attribute-value pair of the final command. There are two types of attribute-level transformation, which are: ❶ **Attribute value update:** we change the value of an attribute in the PTC. The target attributes, for which the values are revised, include the length of a message, the protocol version, algorithms to be used, sequence numbers, and others. We also extract a set of possible values for each attribute from the LLM and generate TCs based on the set for each attribute. For example, we utilize the LLM to make a possible set for the encryption algorithm in the IKE protocol containing values [3–28] and -1 because the former is specified in the specification, and the latter is an exception. ❷ **Attribute drop:** we drop a specific attribute from the message. The attributes to be dropped include all the attributes in the message, such as the algorithms to be used or the protocol version. For example, we drop the encryption algorithm field in the IKE_SA_INIT response message.

**VoWiFi UE testing.** The ATCs generated from the first step are fed into the CONTROLLER to evaluate the security of the VoWiFi UE implementations. The CONTROLLER takes a set of ATCs and processes them one by one. According to each ATC, the CONTROLLER controls AGENTS to send messages through the command-report protocol. The logs are finally collected on the CONTROLLER.

**Decision oracles.** To automatically determine UE misbehavior, we introduce two decision oracles that work over the collected logs and analyze whether the UE's behavior is far from what we expect. These oracles are: ❶ **Function oracle:** it raises an alert if a UE responds to an adversarial message. Because of our highest emphasis on security, we assume that if a UE receives an adversarial message, it should not respond to them. There are three behaviors that a UE can perform when the ATC is executed: (1) ignore the adversarial message, (2) respond with the positive message, that is, the message with which the UE usually replies to the non-adversarial message, and (3) respond with the negative message, that is, the message indicating an error. In case of (1), the oracle determines that the UE behaves correctly; thus, it does not raise an alert. When the UE responds with a positive message (the case (2)), the oracle raises an alert as it means that the UE violates the tested property. Finally, when the UE responds with a negative message (the case (3)), the oracle also raises an alert because there is the possibility of incorrect behavior. ❷ **Liveness oracle:** it checks whether a UE can re-execute the VoWiFi protocol after it receives an adversarial message and disconnects the corresponding session. We expect that the adversarial message should not affect the other sessions. Therefore, the oracle raises an alert if the UE does not send the IKE_SA_INIT request message to an ePDG after the aborted session.

With the above two oracles, we design the testing for each test-case as follows:

(1) **Running an** ATC: VWAttacker performs an adversarial testing based on an ATC for a specific UE.

(2) **Checking an anomalous message flow:** The function oracle checks the logs and raises an alert if it sees an unexpected message from the UE.

(3) **Running a normal registration:** VWAttacker tries to run the normal VoWiFi execution with the UE.

(4) **Checking liveness:** The liveness oracle checks if the UE executes the VoWiFi protocol correctly.

To resolve any non-determinism due to over-the-air testing, this process is repeated three times for each ATCs to get stable results. Note that the execution of the normal registration with the liveness oracle is necessary not only to simply check whether the UE is alive but also to understand the result of the function oracle in more detail. The reason is that we assume that a UE should not respond to an adversarial message. However, we cannot determine whether the lack of response by the UE is due to a crash of the UE, which would represent a false negative of the function oracle. Therefore, we use the liveness oracle to catch the case of the crash.

Finally, the logs on each ATC are labeled with the flags from the decision oracles. We manually analyze the results to pinpoint the issues and the root causes. We discuss the concrete issues in different UEs in Section 7.

## 6 IMPLEMENTATION

This section describes how we implement VWAttacker. The implementation contains several modules written in different languages (see Table 7).

**VoWiFi networking testbed.** We implement the VoWiFi networking testbed on one general purpose machine (e.g., a laptop). As the VoWiFi protocol begins with the ePDG discovery leveraging the DNS protocol after attaching a UE to a WiFi AP, we introduce our controlled WiFi AP and DNS. In detail, we run `hostapd` to make our machine a WiFi AP, and `dnsmasq`, a simple DHCP/DNS server, to assign the private IP address to a UE and to generate a DNS reply that contains the IP address of our ePDG. To this end, we insert one DNS record that maps the name of the ePDG to our IP address of the ePDG into the file that `dnsmasq` refers to. With this setting, we can make a UE to connect to our ePDG since the UE receives a DNS reply containing our ePDG's IP address when a UE attaches to the WiFi AP and sends a DNS query. For an ePDG, an IMS, and an HSS, we use *StrongSwan* [55] version 5.4.5 for ePDG, *Kamailio* [33] version 1.2.3 for IMS, and *FHoSS* [26] for HSS. The ePDG maintains the mapping between the IMSI and key materials (i.e., a UE's secret key and an operator's key) as well as the mapping between the IKE configuration of the P-CSCF and the IP address of the P-CSCF. The former mapping is used to authenticate a UE when establishing the IKE channel, while the latter mapping is used to forward the SIP messages from a UE to the IMS components. As IMS consists of P-CSCF, I-CSCF, and S-CSCF, we run three *Kamailio* processes with different settings. *FHoSS* maintains the mappings between IMSIs and other keys.

**Control architecture.** We implement the CONTROLLER and AGENTS over the VoWiFi networking testbed. The CONTROLLER implementation contains the configurator that sets up the parameters (e.g., the logging directory) required to run the framework, the processor that processes testcases, the logger that maintains logs, and the oracles that label the logs. The UEAGENT implementation includes the initializer that recognizes an attached UE and makes a UE ready (e.g., configuring VoWiFi) through adb. The ePDGAGENT implementation consists of the communicator that receives the commands and reports the results and the processor that manipulates the IKE messages according to the testcase. Similar to the ePDGAGENT, the IMSAGENT implementation contains the communicator and the processor.

**LLM-based property extractor.** We iterate through all 11 specification documents including 3GPP documents [3–5], RFC documents [9, 29, 31, 32, 34, 36, 37, 48, 53], and the conformance test suite [6], and split the documents into smaller chunks. Then, the chunks are embedded using *HuggingFaceEmbeddings* and indexed in a FAISS [21] database for efficient retrieval. We configure a retriever based on this database to search for documents similar to a given query. It uses the FAISS index to find the most similar documents.

Next, we create a text generation pipeline with *HuggingFacePipeline* using an LLM and a tokenizer. We use Mistral-7B-v0.1 [30], Llama-3-8B-Instruct and Llama-3.1-8B-Instruct [8] as text generation models. With a generative model, the framework handles text generation based on the provided input. Finally, we create the RAG pipeline using LangChain [15], where the retriever fetches relevant specification documents from the vector database and the LLM generates properties based on the retrieved specifications and the provided instructions and example properties in the prompt. We implement the property extractor over a Google Colab environment with an NVIDIA L4 GPU with 22.5 GB of GPU RAM, 53 GB of system RAM, and 200 GB of disk space. The models take around one compute hour to generate all the properties. A subset of the LLM-generated properties are listed in Table 2.

**Property-to-testcase encoder.** We write a script that takes the LLM-generated properties and the protocol message flow as input, and leverages the RAG-based LLM model to output the PTCs. We build the protocol message flow based on the diagrams on page 35 of [5] for the UE authentication/authorization with the ePDG, and page 5 of [31] for the UE registration on IMS, to determine the state required to execute the property. Then, we abstract the diagrams into a graph by using NetworkX [1]. For each property, the encoder generates a series of commands by mapping the message name to cause a UE to reach a specific state related to the property, as outlined in the protocol message flow. In case the message name is not in the property, it is taken as input. Then, the encoder appends the message mentioned in the property as the message of the final command in the sequence, thus completing the testcase. In each step, the encoder utilizes the LLM to extract the field names and values of the messages, if not specified in the property.

**Testcase transformer.** We implement the transformer as a Python script, that performs two transformations. The transformer takes PTCs, IKE/SIP error messages, and the list of target attributes and their possible values. After running the script, we finally create 1,116 ATCs.

# 7 EVALUATION

To evaluate the performance of VWATTACKER, we aim to answer the following research questions:

- **RQ1.** How faithful and relevant are the properties generated by TESTGEN?
- **RQ2.** How effective is VWATTACKER in finding issues and attacks on diverse VoWiFi UE implementations?
- **RQ3.** How effective is VWATTACKER, compared to the previous IKE or SIP implementation testing approaches, and the conformance test suite from the VoWiFi standards [6]?

## 7.1 Performance Evaluation of TESTGEN

We evaluate TESTGEN in two dimensions – (1) generation performance and (2) retrieval performance – and report the results (see Table 3).

**Generation Performance.** The generation performance evaluates the quality of properties that a model produces. To evaluate generation performance, we use the metrics listed below: ① **Answer relevancy (AR):** it is used to assess whether the LLM returns concise answers by determining the proportion of sentences in the

LLM output that are relevant to the input. ② **Hallucination rate (HR):** it measures the proportion of hallucinated sentences in an LLM output as this is critical when dealing with LLM-generated outputs. ③ **Bilingual evaluation understudy (BLEU) score:** it measures the overlap between the generated text and the reference text (i.e., ground truth). ④ **Recall-oriented understudy for gisting evaluation (ROUGE) score:** it measures the overlap of n-grams between the generated text and the reference text. ⑤ **Metric for evaluation of translation with explicit ordering (METEOR) score:** it provides a more nuanced evaluation than BLEU when considering synonyms and stemming.

Our results show that RAG effectively improves the generation performance. Further, the advanced models demonstrate higher scores in generation. Table 3 shows that Llama 3.1 performs best in almost all the metrics discussed above, especially in reducing the HR, which is very important for high-quality text generation. We find adding context retrieved from the specification knowledge base further improves the generation performance.

**Retrieval Performance.** As the property generation largely relies on the retrieved context, we measure the retrieval performance based on the following metrics: ① **Contextual precision (CP):** it measures how accurately the retrieved context aligns with the generated text. This metric is calculated as the ratio of relevant information in the generated text to relevant information in the retrieved context. ② **Contextual recall (CR):** it determines the proportion of sentences in the expected output or ground truth that can be attributed to the retrieved context. ③ **Contextual relevancy (CR):** it is the proportion of sentences in the retrieved context relevant to a given input. To understand generation performance in detail, we measure retrieval performance on the models. We find that the main difference between models in retrieval performance is CP. Thus, our conclusion is that the higher generation performance of the advanced models is mainly due to the higher CPs.

**Choice of Models.** We work exclusively with open-source models to ensure transparency and flexibility. After performing basic sanity checks across various open-source LLMs, we select Mistral-7B-v0.1, Llama-3-8B-Instruct, and Llama-3.1-8B-Instruct for in-depth analysis due to their strong performance and alignment with our requirements. These models demonstrate reliable outputs during preliminary evaluations, making them suitable candidates for further exploration within the RAG framework. As we leverage the RAG approach, TESTGEN does not require fine-tuning the LLMs.

## 7.2 Result of Adversarial Testing

We generate 1,116 ATCs based on 63 security properties extracted from TESTGEN. Examples of the properties are shown in Table 2. We find the 13 issues discovered by ADVTEST on 21 UEs (disclosed issues in Table 4, and the lists of UEs per issues in Table 5).

**Support of Weak Algorithms in IKE (Issues #1 – #5).** In the `IKE_SA_INIT` exchange, *the deprecated algorithms MUST NOT or SHOULD NOT be implemented* [48]. We conduct adversarial testing against the property, seeing if a UE establishes the `IKE_SA` with such weak algorithms.

**ATC and VWATTACKER behavior.** Our transformation generates ATCs by setting the operation to update for the attribute security_-association under the message `IKE_SA_INIT` response. The values for the algorithms are automatically extracted by TestGen referring to the specifications. Finally, the ATC is sent to the attached UE. The function oracle raises alerts when UE responds to the adversarial message, which shows the IKE_SA is established with the algorithm. **Result and analysis.** We analyze the logs to see if there is any IKE_SA established with weak algorithms, which are DES, 3DES (encryption algorithms), AUTH_HMAC_MD5_96 (an integrity algorithm), PRF_HMAC_MD5 (a pseudorandom function), 1536-bit MODP, 1024-bit MODP, 768-bit MODP, 1024-bit MODP with 160-bit prime order, 2048-bit MODP with 224-bit prime order, and 2048-bit MODP with 256-bit prime order. Although 3DES is not officially deprecated in the domain of IKEv2, we set it as a weak algorithm since it is getting deprecated in another domain [32]. We find that 9–15 UEs establish weak IKE_SAs. Note that the results show the possibility of a weak channel that can be established with an incorrectly configured ePDG [23]. Also, the results report the UEs that do not comply with the specifications.

| Mistral | | Llama 3 8B | | Llama 3.1 8B | |
|---|---|---|---|---|---|
| With RAG | Without RAG | With RAG | Without RAG | With RAG | Without RAG |
| The UE takes its own copy of the MSK (Master Session Key) as input to generate the AUTH parameter to authenticate the first IKE_SA_INIT message. | The UE takes its own copy of the Master Session Key (MSK) as input to generate the AUTH parameter to authenticate the first Internet Key Exchange Security Associations (IKE_SA_INIT) message in Vo-WiFi. | The UE takes its own copy of the MSK as input to generate the AUTH parameter to authenticate the first IKE_SA_INIT message. | The UE takes its own copy of the MSK (Master Session Key) as input to generate the AUTH parameter to authenticate the first IKE_SA_INIT message. | The UE takes its own copy of the MSK (Master Session Key) as input to generate the AUTH parameter to authenticate the first IKE_SA_INIT message. | The UE takes its own copy of the MSK as input to generate the AUTH parameter to authenticate the first IKE_SA_INIT message. |
| The UE shall take its own copy of the MSK (Master Session Key) as input to generate the AUTH parameter to authenticate the first IKE_SA_INIT message. | The UE shall take its own copy of the Master Session Key (MSK) as input to generate the AUTH parameter to authenticate the first IKE_SA_INIT message in Vo-WIFI. | The UE shall take its own copy of the MSK as input to generate the AUTH parameter to authenticate the first IKE_SA_INIT message. | The UE shall take its own copy of the MSK as input to generate the AUTH parameter to authenticate the first IKE_SA_INIT message. | The UE shall take its own copy of the MSK as input to generate the AUTH parameter to authenticate the first IKE_SA_INIT message. | The UE shall take its own copy of the MSK as input to generate the AUTH parameter to authenticate the first IKE_SA_INIT message. |

**Table 2: Generated Properties**

| Model | Generation Performance | | | | | | | | | | | | | | Retrieval Performance | | |
|---|---|---|---|---|---|---|---|---|---|---|---|---|---|---|---|---|---|
| | AR↑ | | HR↓ | | BLEU↑ | | METEOR↑ | | ROUGE-L↑ | | | | | | CP↑ | CRec↑ | CRel↑ |
| | | | | | | | | | Precision | | Recall | | F1-score | | | | |
| | R | NR | R | NR | R | NR | R | NR | R | NR | R | NR | R | NR | | | |
| Mistral-7B-v0.1 | 0.83 | 0.68 | 0.30 | 0.30 | 0.42 | 0.18 | 0.67 | **0.43** | 0.51 | **0.53** | 0.51 | 0.37 | 0.57 | 0.39 | 0.62 | 0.51 | 0.50 |
| Llama-3-8B-Instruct | 0.84 | **0.94** | 0.05 | 0.31 | 0.50 | 0.22 | 0.64 | 0.42 | 0.65 | 0.48 | 0.76 | 0.49 | 0.68 | **0.46** | 0.92 | **0.61** | **0.69** |
| Llama-3.1-8B-Instruct | **0.90** | 0.85 | **0.04** | **0.05** | **0.53** | **0.24** | **0.68** | **0.43** | **0.84** | 0.52 | **0.73** | 0.46 | **0.93** | 0.56 | **0.93** | 0.56 | 0.66 |

**Table 3: Evaluation results of TestGen (CP: Contextual Precision, CRec: Contextual Recall, AR: Answer Relevancy, HR: Hallucination Rate, R: RAG, NR: No RAG, ↑: higher is better, ↓: lower is better)**

| No. | Testcase | Result |
|---|---|---|
| | **Support of Weak Algorithms in IKE** | |
| 1 | Modify the value specifying the weak encryption algorithm in the IKE_SA_INIT response | The IKE_SA is established with a weak encryption algorithm (e.g., DES) |
| 2 | Modify the value specifying the weak integrity algorithm in the IKE_SA_INIT response | The IKE_SA is established with a weak integrity algorithm (e.g., AUTH_HMAC_MD5_96) |
| 3 | Modify the value specifying the weak pseudorandom function in the IKE_SA_INIT response | The IKE_SA is established with a weak pseudorandom function (e.g., PRF_HMAC_MD5) |
| 4 | Modify the value specifying the weak DH group and key in the IKE_SA_INIT response | 1) The weak DH key (e.g., 1024-bit MODP) is advertised from UE by default and 2) the IKE_SA is established with the key |
| 5 | Substitute the INVALID_KE_PAYLOAD with a weak DH group for the IKE_SA_INIT response | The IKE_SA is established with a weak DH group (e.g., 1024-bit MODP with 160-bit prime order) |
| | **Support of Weak Algorithms in SIP** | |
| 6 | Modify the SIP authentication algorithm in the 401_unauthorized message | The weak SIP authentication algorithm is run (e.g., MD5) |
| 7 | Modify the encryption algorithm for the SIP messages in the 401_unauthorized message | The weak pair of the encryption/integrity algorithm is advertised from UE and it is selected (e.g., DES/HMAC_MD5) |
| | **Zero DH Key with IKE_SA_INIT** | |
| 8 | Drop the key exchange payload from the IKE_SA_INIT response | The DH shared secret is set to 0, disclosing keys for an attacker to decrypt the first two IKE_AUTH messages |
| | **Nonce Bypass with IKE_SA_INIT** | |
| 9 | Drop the nonce payload from the IKE_SA_INIT response | The nonce value from the ePDG is set to 0 in a UE |
| | **DH Group Downgrade** | |
| 10 | Substitute the INVALID_KE_PAYLOAD with a *weaker* DH group than a UE-advertised group for the IKE_SA_INIT response | The DH group is downgraded from the initially advertised one to the weaker one (e.g., 2048-bit MODP → 1024-bit MODP) |

**Table 4: List of issues disclosed by VWAttacker**

**Support of Weak Algorithms in SIP (Issues #6 – #7).** In the 401_unauthorized message, the SIP authentication algorithms and the encryption/integrity algorithms are negotiated. We conduct adversarial testing to see if UE establishes an SIP session with weak algorithms.

ATC **and VWAttacker behavior.** Our transformation generates ATCs by setting the operation to update the attributes under WWW-Authenticate and Security-Server, respectively. The possible values extracted from specifications are assigned to these attributes. VWAttacker runs the ATCs and the function oracle raise an alert if the algorithm is selected.

**Result and analysis.** We analyze the logs to see if any weak algorithm is selected. The algorithms of our interest include MD5 (SIP

authentication), a pair of DES (encryption algorithms), and HMAC_-MD5_96 (an integrity algorithm). Although these algorithms are not explicitly deprecated, they are controversial [35, 46] in using them due to their weaknesses. There are 2 UEs responding with MD5 when they receive the `401_unauthorized` message with MD5. Among them, we find only one UE supports MD5, while the other simply copies the name of the algorithm from `401_unauthorized` and sends the response message that sets the algorithm name to be the copied name. Interestingly, we find that all the UEs advertise the use of DES as an encryption algorithm together with HMAC_MD5_96 as an integrity algorithm. Note that this result demonstrates the impact of a unified testbed. Without completing the IKEv2 protocol, the implementation does not reach this state to trigger this SIP issue.

**Zero DH Key with `IKE_SA_INIT` (Issue #8).** In the initial exchange (i.e., `IKE_SA_INIT`), The `IKE_SA_INIT` response message must contain the key exchange payload that contains the responder's DH key [34]. We check how a UE responds to the message against the property.

ATC **and VWATTACKER behavior.** An ATC is generated by setting the operation to drop for the attribute key_exchange under the message `IKE_SA_INIT` response. When VWATTACKER receives the ATC, the ePDGAgent removes the key exchange payload from the generated `IKE_SA_INIT` response message and sends it to the attached UE.

**Result and analysis.** The function oracle raises alerts in the experiment logs of 8 UEs as the UEs respond to this revised message with the encrypted first `IKE_AUTH` request message, which should not happen. We verify that we can decrypt the message with the 0s of the DH shared secret.

**Threat model and attack.** The first `IKE_AUTH` message contains the UE's IMSI which can be used to track a specific UE, an attacker can know the value by decrypting the message on-the-fly. To perform this attack, an attacker needs to know the target's SPI and forge the `IKE_SA_INIT` response message without the key exchange payload. The threat model is practical as `IKE_SA_INIT` has no integrity protection. When an attacker successfully sends the `IKE_SA_INIT` message without the key exchange payload, a UE sets a DH shared secret to 0 and extracts encryption and integrity keys for the IKE channel. Then, a UE sends the first `IKE_AUTH` message encrypted with the keys. An attacker can know a UE's IMSI by decrypting the message. Although previous work reports that the VoWiFi protocol is vulnerable to IMSI catching attack [10], our finding is a more implementation-specific issue that makes the IMSI catching attack easier without leveraging the DNS infrastructure. Through this attack, in the affected implementations, it is possible for an attacker to cause a DH key security bypass and steal devices' IMSI.

**Nonce Bypass with `IKE_SA_INIT` (Issue #9).** In the initial exchange (i.e., `IKE_SA_INIT`), *the `IKE_SA_INIT` response message must contain the nonce payload* [34] to establish the DH shared key and to avoid the replay attack.

ATC **and VWATTACKER behavior.** Our transformation generates an ATC by setting the operation to drop for the attribute nonce under ike_sa_init. The VWATTACKER directs the ePDGAgent to remove the nonce payload from the `IKE_SA_INIT` response message and to send it to the UE.

**Result and analysis.** We find that the 8 UEs respond to this revised message, caught by our function oracle. The UE accepts `IKE_SA_INIT` and sets the nonce value to zero. Therefore, if a man-in-the-middle attacker can send the message with the nonce payload earlier than the sender, the attacker can successfully set the nonce to zero, substantially reducing the entropy of the secure channel, which is undesirable.

**DH Group Downgrade (Issue #10).** According to [34], the `INVALID_KE_PAYLOAD` error message that contains a specific DH group triggers that *the initiator MUST retry the IKE_SA_INIT with the corrected DH group.* We test a UE if it resends a key on a weaker, thus downgraded, DH group.

ATC **and VWATTACKER behavior.** Our transformation generates an ATC by setting the operation to substitute for the message `IKE_SA_INIT` response and update for the attribute dh_group of the message. The VWATTACKER directs the ePDGAgent to send the `INVALID_KE_PAYLOAD` error response message with a downgraded DH group to the UE.

**Result and analysis.** We find that the 1 − 7 UEs respond to this error message, caught by our function oracle. We see the logs where one UE responds with an arbitrary DH group even if it does not advertise the group. For instance, we can let the UE to send the DH key over the 768-bit MODP group.

**Threat model and attack.** Based on this issue, an attacker can let a UE to establish the weaker IKE channel. To perform this attack, an attacker needs to be the man-in-the-middle and send the adversarial `INVALID_KE_PAYLOAD` specified with the weaker DH group to a UE. Concurrently, this attack is also reported in [24] but on different devices. After a weaker IKE channel is established with the affected implementations, an attacker can decrypt the messages on the downgraded channel using known attacks [24, 41]

## 7.3 Comparison with Previous Work

We compare the capability of VWATTACKER in terms of 1) the ATC generation and 2) the decision oracles in identifying issues with those of other approaches [6, 19, 52, 56, 58] for IKE, SIP, or VoWiFi implementations (see Table 1). As the conformance test suite [6] only checks UE's operational behavior, it cannot detect any issues. Cui *et al.* [19] is unable to detect any of the issues detected by VWATTACKER. This is due to several reasons: first, since [19] only focuses on the IKE protocol, it cannot generate any ATCs related to SIP. Second, it cannot create ATCs #5 and #8 because it cannot replace one message with another. This shows why the message-level transformation is useful. Because SECFUZZ [56] can only automatically detect memory corruption in determining issues, it cannot identify any of issues among our findings. NSFUZZ [52] can only generate ATCs #6 and #7 as it evaluates open-source implementations, including *Kamailio*, by sending messages with random values for specific attributes, but is not able to detect due to the lack of automated detection using the oracles. Note that c07-sip [58] cannot identify any issues because it only tests with specific attributes that are not related to our findings.

## 8 RELATED WORK

Baek *et al.* [10] show that a UE's IMSI can be exposed to an attacker by making a UE connected to a fake IPSec server through a DNS spoofing attack. Also, they show that an attacker can make a UE detaching from a WiFi AP through a WiFi deauthentication frame, which makes WiFi calls unavailable. Xie *et al.* [60] show that there is no defense mechanism in the VoWiFi specification to prevent a UE from associating with insecure networks through an ARP spoofing attack. In addition, they demonstrate that an attacker can get information about events between a UE and the ePDG through traffic analysis. Lu *et al.* [44] demonstrate an availability attack through the IMS messages. By sending fake IMS call messages from a malicious UE, a target UE cannot make a call. Shi *et al.* [54] demonstrate that UEs can be enforced to assign the IMS server's IP address on their interfaces. In addition, they show that there is no restriction on the source of IMS signalling; thus, it can provide fabricated SMS messages to block messages. Gegenhuber *et al.* [23] show that some UEs are vulnerable to the DH downgrade attack and also uncover that more than half of the ePDGs in practice support weak DH groups. Lee *et al.* [40] formally verify the VoWiFi standards. Compared with all these works, VWAttacker is the only systematic and complete framework dealing with the analysis of commercial VoWiFi implementations.

| Issue | # of UEs | Affected UEs |
|---|---|---|
| 1 | 9 (DES) | HTC U11 life, LG Stylo 6, Moto e5 plus, Nokia G100, OnePlus 9R, OnePlus Nord 20, Pixel 4a, UMIDIGI A13 Pro, ZTE Stage 5G |
| | 15 (3DES) | BlackCyber I14, BlackCyber I15, Blackview A55, HTC U11 life, LG Stylo 6, Moto e5 plus, Nokia G100, NUU B15, OnePlus 9R, OnePlus Nord 20, Pixel 4a, TCL 40XL, Ulefone Note 14, UMIDIGI A13 Pro, ZTE Stage 5G |
| 2 | 15 (HMAC_MD5_96) | BlackCyber I14, BlackCyber I15, Blackview A55, HTC U11 life, LG Stylo 6, Moto e5 plus, Nokia G100, NUU B15, OnePlus 9R, OnePlus Nord 20, Pixel 4a, TCL 40XL, Ulefone Note 14, UMIDIGI A13 Pro, ZTE Stage 5G |
| 3 | 6 (PRF_HMAC_MD5) | BlackCyber I14, BlackCyber I15, Blackview A55, NUU B15, TCL 40XL, Ulefone Note 14 |
| 4 | 12 (1024-bit MODP) | BlackCyber I14, BlackCyber I15, Blackview A55, NUU B15, TCL 40XL, Ulefone Note 14 |
| 5 | 19 (1024-bit MODP) | HTC U11 life, LG Stylo 6, Moto e5 plus, Nokia G100, OnePlus 9R, OnePlus Nord 20, Pixel 4a, UMIDIGI A13 Pro, ZTE Stage 5G |
| | 15 (1536-bit MODP) | BlackCyber I14, BlackCyber I15, Blackview A55, HTC U11 life, LG Stylo 6, Moto e5 plus, Nokia G100, NUU B15, OnePlus 9R, OnePlus Nord 20, Pixel 4a, TCL 40XL, Ulefone Note 14, UMIDIGI A13 Pro, ZTE Stage 5G |
| | 6 (1024-bit MODP 160-bit prime order) | BlackCyber I14, BlackCyber I15, Blackview A55, NUU B15, TCL 40XL, Ulefone Note 14 |
| | 6 (2048-bit MODP 224-bit prime order) | BlackCyber I14, BlackCyber I15, Blackview A55, NUU B15, TCL 40XL, Ulefone Note 14 |
| | 6 (2048-bit MODP 256-bit prime order) | BlackCyber I14, BlackCyber I15, Blackview A55, NUU B15, TCL 40XL, Ulefone Note 14 |
| 6 | 2 (MD5) | LG Stylo 6, Pixel 6a |
| 7 | 21 (DES/HMAC_MD5_96) | BlackCyber I14, BlackCyber I15, Blackview A55, HTC U11 life, LG Stylo 6, Moto e5 plus, Nokia G100, NUU B15, OnePlus 9R, OnePlus Nord 20, Pixel 4a, Pixel 6a, Samsung Galaxy A21s, Samsung Galaxy A34 5G, Samsung Galaxy S6, TCL 40XL, Ulefone Note 14, UMIDIGI A13 Pro, ZTE Stage 5G |
| 8 | 8 (No key exchange payload) | BlackCyber I14, HTC U11 life, Moto e5 plus, Nokia G100, OnePlus 9R, OnePlus Nord 20, Pixel 4a, ZTE Stage 5G |
| 9 | 8 (No nonce payload) | BlackCyber I14, HTC U11 life, Moto e5 plus, Nokia G100, OnePlus 9R, OnePlus Nord 20, Pixel 4a, ZTE Stage 5G |
| 10 | 1 (2048-bit MODP → 768-bit MODP ) | LG Stylo 6 |
| | 7 (2048-bit MODP → 1024-bit MODP ) | HTC U11 life, LG Stylo 6, Mote e5 plus, Nokia G100, OnePlus 9R, OnePlus Nord 20 |
| | 7 (2048-bit MODP → 1536-bit MODP ) | LG Stylo 6 |
| | 1 (2048-bit MODP → 1024-bit MODP 160-bit prime order) | LG Stylo 6 |
| | 1 (2048-bit MODP → 2048-bit MODP 224-bit prime order) | LG Stylo 6 |
| | 1 (2048-bit MODP → 2048-bit MODP 256-bit prime order) | LG Stylo 6 |
| 11 | 1 (reboot) | LG Stylo 6 |
| | 7 (WiFi reset) | HTC U11 life, Nokia G100, OnePlus 9R, OnePlus Nord 20, Pixel 4a, Samsung Galaxy S6, Ulefone Note 14 |
| 12 | 1 (reboot) | LG Stylo 6 |
| | 7 (WiFi reset) | HTC U11 life, Nokia G100, OnePlus 9R, OnePlus Nord 20, Pixel 4a, Samsung Galaxy S6, Ulefone Note 14 |
| 13 | 1 (reboot) | LG Stylo 6 |
| | 7 (WiFi reset) | HTC U11 life, Nokia G100, OnePlus 9R, OnePlus Nord 20, Pixel 4a, Samsung Galaxy S6, Ulefone Note 14 |

Table 5: Mapping between the issues and the UEs

| Issue (ATC no.) | VWATTACKER | Conformance test suite [6] | Cui et al. [19] | SECFUZZ [56] | NSFUZZ [52] | α07-ssp [58] |
|---|---|---|---|---|---|---|
| 1 | ● | - | - | - | - | - |
| 2 | ● | - | - | - | - | - |
| 3 | ● | - | - | - | - | - |
| 4 | ● | - | - | - | - | - |
| 5 | ● | - | - | - | - | - |
| 6 | ● | - | - | - | - | - |
| 7 | ● | - | - | - | - | - |
| 8 | ● | - | - | - | - | - |
| 9 | ● | - | - | - | - | - |

Table 6: Comparison VWAttacker with other approaches

## 9 CONCLUSION

In this paper, we design and implement the systematic and automated framework called VWAttacker to analyze the security of VoWiFi UEs. It consists of three parts – the LLM-based testcase generation (TestGen), the adversarial testing (AdvTest), and the result analysis. With VWAttacker, we uncover 13 issues on VoWiFi UE and show our better capabilities over other approaches in generating ATCs and determining issues.

## REFERENCES

[1] [n.d.]. *networkx.* https://networkx.org/.
[2] 2019. Voice over WiFi (VoWiFi) Market Size & Share Analysis - Growth Trends & Forecasts (2025 - 2030). https://www.mordorintelligence.com/industry-reports/voice-over-wifi-vowifi-market. (Accessed on 05/27/2025).
[3] 3GPP TS 23.402. 2019. Architecture enhancements for non-3GPP accesses Release 16. https://portal.3gpp.org/desktopmodules/Specifications/SpecificationDetails.aspx?specificationId=850. (Accessed on 09/04/2024).
[4] 3GPP TS 24.301. 2019. Non-Access-Stratum (NAS) protocol for Evolved Packet System (EPS); Stage 3. https://portal.3gpp.org/desktopmodules/Specifications/SpecificationDetails.aspx?specificationId=1072. (Accessed on 09/04/2024).
[5] 3GPP TS 33.402. 2020. Security aspects of non-3GPP accesses Release 16. https://portal.3gpp.org/desktopmodules/Specifications/SpecificationDetails.aspx?specificationId=2297. (Accessed on 09/04/2024).
[6] 3GPP TS 36.523-1. 2021. Evolved Universal Terrestrial Radio Access (E-UTRA) and Evolved Packet Core (EPC); User Equipment (UE) conformance specification; Part 1: Protocol conformance specification Release 16. https://www.etsi.org/deliver/etsi_ts/136500_136599/13652301/16.08.00_60/ts_13652301v160800p.pdf. (Accessed on 09/04/2024).
[7] Mayank Agarwal, Santosh Biswas, and Sukumar Nandi. 2018. An efficient scheme to detect evil twin rogue access point attack in 802.11 Wi-Fi networks. *International Journal of Wireless Information Networks* 25, 2 (2018), 130–145.
[8] AI@Meta. 2024. Llama 3 Model Card. (2024). https://github.com/meta-llama/llama3/blob/main/MODEL_CARD.md

[9] J Arkko and H Haverinen. 2006. Extensible Authentication Protocol Method for 3rd Generation Authentication and Key Agreement (EAP-AKA). https://datatracker.ietf.org/doc/html/rfc4187. (Accessed on 09/05/2024).

[10] Jaejong Baek, Sukwha Kyung, Haehyun Cho, Ziming Zhao, Yan Shoshitaishvili, Adam Doupé, and Gail-Joon Ahn. 2018. Wi not calling: Practical privacy and availability Attacks in Wi-Fi calling. In *Proceedings of the 34th Annual Computer Security Applications Conference*. 278–288.

[11] David Basin, Jannik Dreier, Lucca Hirschi, Saša Radomirovic, Ralf Sasse, and Vincent Stettler. 2018. A formal analysis of 5G authentication. In *Proceedings of the 2018 ACM SIGSAC conference on computer and communications security*. 1383–1396.

[12] Scott Bradner. 1997. RFC2119: Key words for use in RFCs to Indicate Requirement Levels. https://datatracker.ietf.org/doc/html/rfc2119. (Accessed on 10/18/2023).

[13] Scott Bradner. 1997. RFC2119: Key words for use in RFCs to Indicate Requirement Levels. https://datatracker.ietf.org/doc/html/rfc2119. (Accessed on 10/18/2023).

[14] Scott Bradner. 1997. RFC2119: Key words for use in RFCs to Indicate Requirement Levels. https://datatracker.ietf.org/doc/html/rfc2119. (Accessed on 09/06/2024).

[15] Harrison Chase. 2022. LangChain. https://github.com/langchain-ai/langchain

[16] Yi Chen, Yepeng Yao, XiaoFeng Wang, Dandan Xu, Chang Yue, Xiaozhong Liu, Kai Chen, Haixu Tang, and Baoxu Liu. 2021. Bookworm game: Automatic discovery of the vulnerabilities through documentation analysis. In *2021 IEEE Symposium on Security and Privacy (SP)*. IEEE, 1197–1214.

[17] Cisco. 2019. Cisco Visual Networking Index: Global Mobile Data Traffic Forecast Update, 2017-2022. https://s3.amazonaws.com/media.mediapost.com/uploads/CiscoForecast.pdf. (Accessed on 09/06/2024).

[18] Cas Cremers, Benjamin Kiesl, and Niklas Medinger. 2020. A Formal Analysis of IEEE 802.11's WPA2: Countering the Kracks Caused by Cracking the Counters. In *29th USENIX Security Symposium (USENIX Security 20)*. USENIX Association, 1–17. https://www.usenix.org/conference/usenixsecurity20/presentation/cremers

[19] Yanpeng Cui, Ting Yu, and Jianwei Hu. 2018. IKEv2 Protocol Fuzzing Test on Simulated ASA. In *2018 IEEE International Conference on Smart Internet of Things (SmartIoT)*. IEEE, 111–116.

[20] Danny Dolev and Andrew C. Yao. 1983. On the security of public key protocols. *IEEE Transactions on information theory* 29, 2 (1983), 198–208.

[21] Matthijs Douze, Alexandr Guzhva, Chengqi Deng, Jeff Johnson, Gergely Szilvasy, Pierre-Emmanuel Mazaré, Maria Lomeli, Lucas Hosseini, and Hervé Jégou. 2024. The Faiss library. (2024). arXiv:2401.08281 [cs.LG]

[22] Matheus E Garbelini, Zewen Shang, Sudipta Chattopadhyay, Sumei Sun, and Ernest Kurniawan. 2022. Towards automated fuzzing of 4g/5g protocol implementations over the air. In *GLOBECOM 2022-2022 IEEE Global Communications Conference*. IEEE, 86–92.

[23] Gabriel K Gegenhuber, Florian Holzbauer, Philipp É Frenzel, Edgar Weippl, and Adrian Dabrowski. 2024. Diffie-Hellman Picture Show: Key Exchange Stories from Commercial {VoWiFi} Deployments. In *33rd USENIX Security Symposium (USENIX Security 24)*. 451–468.

[24] Gabriel K Gegenhuber, Florian Holzbauer, Philipp É. Frenzel, Edgar Weippl, and Adrian Dabrowski. 2024. Diffie-Hellman Picture Show: Key Exchange Stories from Commercial VoWiFi Deployments. In *33rd USENIX Security Symposium (USENIX Security 24)*. USENIX Association, Philadelphia, PA, 451–468. https://www.usenix.org/conference/usenixsecurity24/presentation/gegenhuber

[25] GSMA. 2024. *The Mobile Economy*. Technical Report. GSMA.

[26] Supreeth Herle. 2022. FHoSS Github Site. https://github.com/herlesupreeth/FHoSS. (Accessed on 09/05/2024).

[27] Syed Rafiul Hussain, Mitziu Echeverria, Imtiaz Karim, Omar Chowdhury, and Elisa Bertino. 2019. 5GReasoner: A property-directed security and privacy analysis framework for 5G cellular network protocol. In *Proceedings of the 2019 ACM SIGSAC Conference on Computer and Communications Security*. 669–684.

[28] Syed Rafiul Hussain, Imtiaz Karim, Abdullah Al Ishtiaq, Omar Chowdhury, and Elisa Bertino. 2021. Noncompliance as deviant behavior: An automated black-box noncompliance checker for 4g lte cellular devices. In *Proceedings of the 2021 ACM SIGSAC Conference on Computer and Communications Security*. 1082–1099.

[29] Ari Huttunen, Brian Swander, Victor Volpe, Larry DiBurro, and Markus Stenberg. 2005. UDP encapsulation of IPsec ESP packets. https://datatracker.ietf.org/doc/html/rfc3948. (Accessed on 09/05/2024).

[30] Albert Q. Jiang, Alexandre Sablayrolles, Arthur Mensch, Chris Bamford, Devendra Singh Chaplot, Diego de las Casas, Florian Bressand, Gianna Lengyel, Guillaume Lample, Lucile Saulnier, Lélio Renard Lavaud, Marie-Anne Lachaux, Pierre Stock, Teven Le Scao, Thibaut Lavril, Thomas Wang, Timothée Lacroix, and William El Sayed. 2023. Mistral 7B. arXiv:2310.06825 [cs.CL] https://arxiv.org/abs/2310.06825

[31] A Johnston, S Donovan, R Sparks, C Cunningham, and K Summers. 2003. Session Initiation Protocol (SIP) Basic Call Flow Examples. https://datatracker.ietf.org/doc/html/rfc3665. (Accessed on 09/06/2024).

[32] B. Kaduk and M. Short. 2018. RFC 8429: Deprecate Triple-DES (3DES) and RC4 in Kerberos. https://datatracker.ietf.org/doc/html/rfc8429. (Accessed on 09/06/2024).

[33] Kamailio. 2001. Welcome To Kamailio - The Open Source SIP Server. https://www.kamailio.org/w/. (Accessed on 09/05/2024).

[34] Charlie Kaufman, Paul Hoffman, Yoav Nir, Pasi Eronen, and Tero Kivinen. 2010. Internet key exchange protocol version 2 (IKEv2). https://datatracker.ietf.org/doc/html/rfc5996. (Accessed on 09/05/2024).

[35] S. Kelly. 2006. Security Implications of Using the Data Encryption Standard (DES). https://www.rfc-editor.org/rfc/rfc4772.html. (Accessed on 09/06/2024).

[36] S Kent. 2005. IP Encapsulating Security Payload (ESP). https://datatracker.ietf.org/doc/html/rfc4303. (Accessed on 09/05/2024).

[37] S. Kent and K. Seo. 2005. Security Architecture for the Internet Protocol. https://datatracker.ietf.org/doc/html/rfc4301. (Accessed on 09/06/2024).

[38] Hongil Kim, Dongkwan Kim, Minhee Kwon, Hyungseok Han, Yeongjin Jang, Dongsu Han, Taesoo Kim, and Yongdae Kim. 2015. Breaking and fixing volte: Exploiting hidden data channels and mis-implementations. In *Proceedings of the 22nd ACM SIGSAC Conference on Computer and Communications Security*. 328–339.

[39] Hongil Kim, Jiho Lee, Eunkyu Lee, and Yongdae Kim. 2019. Touching the untouchables: Dynamic security analysis of the LTE control plane. In *2019 IEEE Symposium on Security and Privacy (SP)*. IEEE, 1153–1168.

[40] Hyunwoo Lee, Imtiaz Karim, Ninghui Li, and Elisa Bertino. 2022. VWAnalyzer: A Systematic Security Analysis Framework for the Voice over WiFi Protocol. In *Proceedings of the 2022 ACM on Asia Conference on Computer and Communications Security*. 182–195.

[41] Joonhee Lee, Hyunwoo Lee, Jongheon Jeong, Doowon Kim, and Ted Taekyoung Kwon. 2021. Analyzing Spatial Differences in the TLS Security of Delegated Web Services. In *Proceedings of the 2021 ACM Asia Conference on Computer and Communications Security*. 475–487.

[42] Patrick Lewis, Ethan Perez, Aleksandra Piktus, Fabio Petroni, Vladimir Karpukhin, Naman Goyal, Heinrich Küttler, Mike Lewis, Wen-tau Yih, Tim Rocktäschel, et al. 2020. Retrieval-augmented generation for knowledge-intensive nlp tasks. *Advances in Neural Information Processing Systems* 33 (2020), 9459–9474.

[43] Chi-Yu Li, Guan-Hua Tu, Chunyi Peng, Zengwen Yuan, Yuanjie Li, Songwu Lu, and Xinbing Wang. 2015. Insecurity of voice solution VoLTE in LTE mobile networks. In *Proceedings of the 22nd ACM SIGSAC Conference on Computer and Communications Security*. 316–327.

[44] Yu-Han Lu, Chi-Yu Li, Yao-Yu Li, Sandy Hsin-Yu Hsiao, Tian Xie, Guan-Hua Tu, and Wei-Xun Chen. 2020. Ghost calls from operational 4G call systems: IMS vulnerability, call DoS attack, and countermeasure. In *Proceedings of the 26th Annual International Conference on Mobile Computing and Networking*. 1–14.

[45] Xiaoyue Ma, Lannan Luo, and Qiang Zeng. 2024. From One Thousand Pages of Specification to Unveiling Hidden Bugs: Large Language Model Assisted Fuzzing of Matter {IoT} Devices. In *33rd USENIX Security Symposium (USENIX Security 24)*. 4783–4800.

[46] 3GPP TSG-SA3 Meeting. 2021. Change Request (S3-214307 -r2 1). https://www.3gpp.org/ftp/tsg_sa/WG3_Security/TSGS3_105e/Inbox/Drafts/draft_S3-214307-r2_Securityupdatesforalgorithmsandprotocolsin33.203.docx. (Accessed on 09/06/2024).

[47] Diogo Mónica and Carlos Ribeiro. 2011. Wifihop-mitigating the evil twin attack through multi-hop detection. In *European Symposium on Research in Computer Security*. Springer, 21–39.

[48] Y. Nir, T. Kivinen, P. Wouters, and D. Migault. 2017. Algorithm Implementation Requirements and Usage Guidance for the Internet Key Exchange Protocol Version 2 (IKEv2). https://datatracker.ietf.org/doc/html/rfc8247. (Accessed on 09/06/2024).

[49] Osmocomm. 2024. osmo-ePDG - VoWiFi Evolved Packet Data Gateway. https://osmocom.org/projects/osmo-epdg. (Accessed on 09/06/2024).

[50] CheolJun Park, Sangwook Bae, BeomSeok Oh, Jiho Lee, Eunkyu Lee, Insu Yun, and Yongdae Kim. 2022. DoLTEst: In-depth Downlink Negative Testing Framework for LTE Devices. In *USENIX Security Symposium*.

[51] Chunyi Peng, Chi-Yu Li, Hongyi Wang, Guan-Hua Tu, and Songwu Lu. 2014. Real threats to your data bills: Security loopholes and defenses in mobile data charging. In *Proceedings of the 2014 ACM SIGSAC Conference on Computer and Communications Security*. 727–738.

[52] Shisong Qin, Fan Hu, Bodong Zhao, Tingting Yin, and Zhang. 2022. Registered report: nsfuzz: towards efficient and state-aware network service fuzzing. In *International Fuzzing Workshop (FUZZING)*.

[53] Jonathan Rosenberg, Henning Schulzrinne, Gonzalo Camarillo, Alan Johnston, Jon Peterson, Robert Sparks, Mark Handley, and Eve Schooler. 2002. SIP: Session Initiation Protocol. https://datatracker.ietf.org/doc/html/rfc3261. (Accessed on 09/05/2024).

[54] Jingwen Shi, Sihan Wang, Min-Yue Chen, Guan-Hua Tu, Tian Xie, Man-Hsin Chen, Chi-Yu Li, Chi-Yu Li, and Chunyi Peng. 2024. IMS is Not That Secure on Your 5G/4G Phones. In *Proceedings of the 30th Annual International Conference on Mobile Computing and Networking*. 513–527.

[55] strongSwan. 2005. strongSwan: the OpenSource IPsec-based VPN Solution. https://www.strongswan.org/. (Accessed on 09/05/2024).

```
 1  {"testcases": [
 2  {"testcase": [
 3    {"receiver": "ue",                    Message
 4     "name": "enable_vowifi",
 5     "reporter": "epdg",
 6     "id": "turn-on-wifi"                  Command
 7  ]},
 8    {"receiver": "ue",                    Message
 9     "name": "enable_vowifi",
10     "reporter": "epdg",
11     "id": "update-ealg-null"},            Command
12
13    {"receiver": "epdg",                  Message
14     "name": "ike_sa_init_response",
15     "reporter": "epdg"}                   Command
16    {"receiver": "epdg",                  Message
17     "name": "ike_auth_1_response",
18     "reporter": "epdg"}                   Command
19    {"receiver": "epdg",                  Message
20     "name": "ike_auth_2_response",
21     "reporter": "epdg"}                   Command
22    {"receiver": "epdg",                  Message
23     "name": "ike_auth_3_response",
24     "reporter": "ims"}                    Command
25    {"receiver": "ims",                   Message
26     "name": "401_unauthorized",
27     "reporter": "ims",                    Command
28     "sub": [                             Attribute
29       {"name": "security_server",
30        "sub": [                          Attribute
31          {"name": "ealg",
32           "op": "update",
33           "type": "string",
34           "value": "null"}]}
35  }]}]}
```

**Figure 5: Example of a TC.**

[56] Petar Tsankov, Mohammad Torabi Dashti, and David Basin. 2012. SECFUZZ: Fuzz-testing security protocols. In *2012 7th International Workshop on Automation of Software Test (AST)*. IEEE, 1–7.

[57] Kai Tu, Abdullah Al Ishtiaq, Syed Md Mukit Rashid, Yilu Dong, Weixuan Wang, Tianwei Wu, and Syed Rafiul Hussain. 2024. Logic Gone Astray: A Security Analysis Framework for the Control Plane Protocols of 5G Basebands. In *33rd USENIX Security Symposium (USENIX Security 24)*. USENIX Association, Philadelphia, PA, 3063–3080. https://www.usenix.org/conference/usenixsecurity24/presentation/tu

[58] Christian Wieser, Marko Laakso, and Henning G Schulzrinne. 2003. Security testing of SIP implementations. (2003).

[59] Jianliang Wu, Yuhong Nan, Vireshwar Kumar, Dave (Jing) Tian, Antonio Bianchi, Mathias Payer, and Dongyan Xu. 2020. BLESA: Spoofing Attacks against Reconnections in Bluetooth Low Energy. In *14th USENIX Workshop on Offensive Technologies (WOOT 20)*. USENIX Association. https://www.usenix.org/conference/woot20/presentation/wu

[60] Tian Xie, Guan-Hua Tu, Chi-Yu Li, Chunyi Peng, Jiawei Li, and Mi Zhang. 2018. The dark side of operational Wi-Fi calling services. In *2018 IEEE Conference on Communications and Network Security (CNS)*. IEEE, 1–11.

## A LIST OF UES

Below is a list of UEs that are analyzed by the adversarial testing based on VWATTACKER.

## B LLM DETAILS

### B.1 Prompt

To extract the properties from the specifications we carefully design the prompts for optimal performance. We follow the following steps while designing the prompts. **(1) Clarity and Specificity.** To achieve precise and relevant results, we craft prompts with clear and specific instructions. **(2) Incorporating Context.** We include relevant contextual information retrieved by the retrieval model to enhance the LLM's understanding of the task. **(3) Assigning Roles.** To guide the LLM's responses, we assign it a specific role of a Vo-WiFi expert. This technique proves highly effective in aligning the outputs with the perspective or expertise we seek. **(4) Providing Examples.** We employ "few-shot prompting" by including examples in our prompts. These examples serve as a reference point for the desired format or style, enabling the LLM to better replicate the structure and tone in its responses. **(5) Structuring.** We carefully organize our prompt into distinct sections for instructions, context, and output format. This structured approach helps the LLM parse the information more efficiently and produce well-organized responses. **(6) Refining Through Iteration.** Analyzing and refining prompts is a continuous process. We review the LLM's responses and adjust our prompts to improve clarity, add necessary context, or rephrase instructions. This iterative approach consistently enhances the quality of the results. **(7) Experiment with Techniques** We experiment with different prompting techniques, such as zero-shot, one-shot, and chain-of-thought prompting, to determine the most effective approach for each task. These trials help us identify strategies that work best for specific applications. **(8) Defining Output Format.** We explicitly state the desired output format. By specifying the structure, we ensure the responses are organized in the most useful way for our needs. Below is an example prompt we use to extract properties from the specifications.

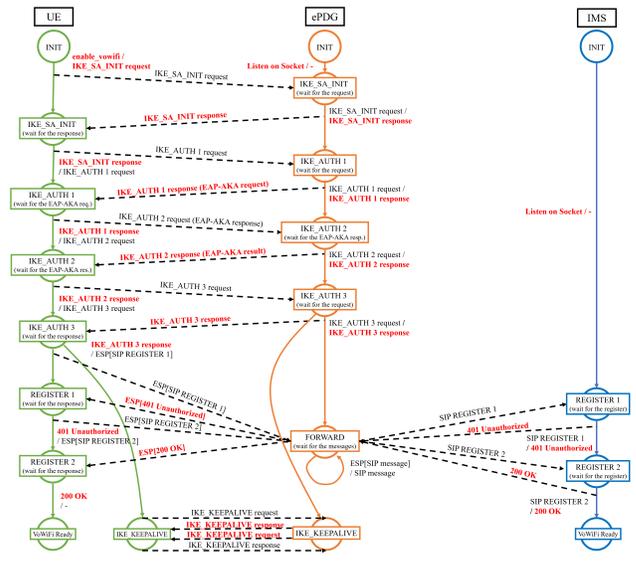

**Figure 6: High-level overview of the protocol flow used for property-to-testcase encoder.**

| Module | Base Implementation | Language | Lines-of-code |
|---|---|---|---|
| **VoWiFi Testbed** | | | |
| CONTROLLER | - | Java | 2783 |
| UEAGENT | - | Python | 442 |
| ePDGAGENT | StrongSwan | C | 1789 |
| IMSAGENT | Kamailio | C | 1195 |
| **Testcase Generation (TESTGEN)** | | | |
| EXTRACTOR | - | Python | 212 |
| ENCODER | - | Python | 96 |
| **Adversarial Testing (ADVTEST)** | | | |
| TRANSFORMER | - | Python | 350 |
| ORACLES | - | Java | 143 |
| **Result Analysis** | | | |
| ANALYSIS | - | Python | 135 |

**Table 7: Implementation detail of VWATTACKER**

| No. | Device Vendor | Device Model | Android Version | Baseband Vendor | Baseband Model |
|-----|---------------|--------------|-----------------|-----------------|----------------|
| 1 | Blackcyber | I14 Pro Max | 13 | Qualcomm | Snapdragon 888 |
| 2 | Blackcyber | I15 Pro Max | 13 | Qualcomm | Snapdragon 8 Gen2 |
| 3 | Blackview | A55 | 11 | MediaTek | Helio A22 (MT6761) |
| 4 | Google | Pixel 4a | 13 | Qualcomm | Snapdragon 765G |
| 5 | Google | Pixel 6a | 12 | Google | Tensor |
| 6 | HTC | U11 life | 8 | Qualcomm | Snapdragon 630 |
| 7 | LG | Stylo 6 (LM-Q730TM) | 10 | MediaTek | Helio P35 (MT6765) |
| 8 | Motorola | Moto $E^5$ Plus | 8 | Qualcomm | Snapdragon 430 |
| 9 | Nokia | G100 | 12 | Qualcomm | Snapdragon 665 |
| 10 | NUU | B15 (S6701L) | 11 | MediaTek | Helio G80 (MT6768) |
| 11 | OnePlus | Nord N20 (CPH2459) | 12 | Qualcomm | Snapdragon 695 |
| 12 | OnePlus | 9R | 11 | Qualcomm | Snapdragon 870 |
| 13 | Samsung | Galaxy S6 (G920T) | 7 | Samsung | Exynos 7 Octa 7420 |
| 14 | Samsung | Galaxy A21s | 12 | Samsung | Exynos 850 |
| 15 | Samsung | Galaxy A34 5G | 14 | MediaTek | Dimensity 1080 (MT6877V/TTZA) |
| 16 | Samsung | Galaxy A35 5G | 14 | Samsung | Exynos 1380 |
| 17 | Samsung | Galaxy A04 | 14 | MediaTek | Helio P35 (MT6765V/CB) |
| 18 | TCL | 40XL (T608M) | 13 | MediaTek | Helio (MT6765V/CA) |
| 19 | Ulefone | Note 14 | 12 | MediaTek | Helio A22 |
| 20 | UMIDIGI | A13 Pro (MP05) | 11 | Unisoc | T610 |
| 21 | ZTE | Stage 5G (A2020N3) | 9 | Qualcomm | Snapdragon 855 |

Table 8: List of UEs supported by VWATTACKER (Ordered by device vendors).

```
<|system|>
% Role Declaration
You are a Vo-WiFi expert. Your task is to extract
    properties of Vo-WiFi from the contexts given from
    specifications.

Below you will find the basic structure of Properties in
    a Vo-WiFi specification.

% Rules
Properties often
1. Use "shall", "must", or "should" to indicate mandatory
    actions or strong recommendations.
2. Describe specific actions, such as taking input,
    sending parameters, or generating output.
3. Include terms like Input, Output, Parameter etc.
4. Specify interactions between network components (e.g.,
    UE, ePDG, AAA Server) and the data exchanged (e.g.,
    AUTH parameter, Notify payload).
5. Describe steps in a process and the dependencies
    between them, specify conditions or contexts for
    actions to occur.

% Instructions
1. Be concise while generating; only give the extracted
    properties as a response, and don't add anything on
    your own.

% Example Properties
Some example properties are:
1. The UE shall take its own copy of the MSK as input to
    generate the AUTH parameter to authenticate the
    first IKE_SA_INIT message.
```

```
2. The AUTH parameter is sent to the ePDG. The UE
    includes a Notify payload ANOTHER_AUTH_FOLLOWS
    indicating to the ePDG that another authentication
    and authorization round will follow.
3. The UE sends the identity in the private network in
    IDi payload that is used for the next authentication
    and authorization with the External AAA Server and
    without an AUTH payload.

% Context Block
You will find the required information about vo-wifi
    properties in the following context:

{context} % retrieved from the specifications

<|assistant|>  % Assistant Output Section
```

| Sl | Property |
|---|---|
| 1 | The UE shall take its own copy of the MSK as input to generate the AUTH parameter to authenticate the first IKE_SA_INIT message. |
| 2 | The UE takes its own copy of the MSK (Master Session Key) as input to generate the AUTH parameter to authenticate the first IKE_SA_INIT message. |
| 3 | The AUTH parameter is sent to the ePDG, and the UE includes a Notify payload ANOTHER_AUTH_FOLLOWS to indicate that another authentication and authorization round will follow. |
| 4 | The UE sends its identity in the private network in the IDi payload for the next authentication and authorization with the External AAA Server and without an AUTH payload. |
| 5 | The UE shall take its own copy of the MSK as input to generate the AUTH parameter to authenticate the first IKE_SA_INIT message. |
| 6 | EAP-AKA, as specified in RFC 4187, within IKEv2, as specified in RFC 5996, shall be used to authenticate UEs, and certificates used for authentication of the ePDG shall meet the certificate profiles given in TS 33.310. |
| 7 | The ePDG shall authenticate itself to the UE with an identity that is the same as the FQDN of the ePDG determined by the ePDG selection procedures defined in TS 23.402, and this identity shall be contained in the IKEv2 ID_FQDN payload and shall match a dNSName SubjectAltName component in the ePDG's certificate. |
| 8 | The UE shall use the Configuration Payload of IKEv2 to obtain the Remote IP address. |
| 9 | Replay protection is provided in IKEv2 as the UE and ePDG generate nonces as input to derive the encryption and authentication keys, preventing intermediate nodes from modifying or changing the user identity. |
| 10 | The UE omits the AUTH parameter in order to indicate to the ePDG that it wants to use EAP over IKEv2. |
| 11 | When the UE requests with a CERTREQ payload, the ePDG responds by sending the certificates requested by the UE in the CERT payload. To protect the previous message in the IKE_SA_INIT exchange, the ePDG includes an AUTH payload in the response. |
| 12 | The UE checks the authentication parameters and responds to the authentication challenge, and the IKE_AUTH request message includes the EAP message (EAP-Response/AKA-Challenge) containing the UE's response to the authentication challenge. |
| 13 | The UE takes its own copy of the MSK as input to generate the AUTH parameter, and the AUTH parameter is sent to the ePDG. |
| 14 | The UE shall send X.509 certificate - Signature payloads with encoding value 4. |
| 15 | The UE shall not assume that any except the first IKEv2 CERT payload is ordered in any way. |
| 16 | The UE shall be able to support certificate paths containing up to four certificates, where the intermediate CA certificates and the ePDG certificate are obtained from the IKEv2 CERT payload and the self-signed CA certificate is obtained from a UE local store of trusted root certificates. |
| 17 | The UE shall be prepared to receive irrelevant certificates, or certificates it does not understand. |
| 18 | The UE shall be able to process certificates even if naming attributes are unknown. |
| 19 | The UE shall support both UTCTime and GeneralizedTime encoding for validity time. |
| 20 | The UE shall check the validity time, and reject certificates that are either not yet valid or are expired. |
| 21 | The UE shall support processing of the BasicConstraints, NameConstraints, and KeyUsage extensions. |
| 22 | Support for OCSP is mandatory in the UE. |
| 23 | The UE should send an OCSP request message to the OCSP server after the tunnel is established, and before user data is transmitted, to check the certificate status of the ePDG. |
| 24 | The UE shall establish a new IPsec tunnel with the new ePDG as described in subclause 8.2.2. |
| 25 | The UE receives an IKE_AUTH Response message from the ePDG, containing its identity, a certificate, and the AUTH parameter to protect the previous message it sent to the UE. |
| 26 | The UE shall re-establish the IPsec Tunnel for the corresponding PDN connection after its release. |
| 27 | The first certificate provided MUST contain the public key used to verify the AUTH field. |
| 28 | The responder might use some other IDr to finish the exchange. |
| 29 | If the initiator guesses the wrong Diffie-Hellman group during the IKE_SA_INIT, it must retry the IKE_SA_INIT with the corrected Diffie-Hellman group, and it should again propose its full supported set of groups, while picking an element of the selected group for its KE value. |
| 30 | The IKE SA is still created as usual, and the Notify message types that do not prevent an IKE SA from being set up include at least NO_PROPOSAL_CHOSEN, TS_UNACCEPTABLE, SINGLE_PAIR_REQUIRED, INTERNAL_ADDRESS_FAILURE, and FAILED_CP_REQUIRED. |
| 31 | If the failure is related to creating the IKE SA, the IKE SA is not created. The information needs to be treated with caution, assuming the peer receiving the Notify error message has not yet authenticated the other end, or if the peer fails to authenticate the other end for some reason. |
| 32 | The responder MUST reject the request and indicate its preferred Diffie-Hellman group in the response. |
| 33 | INFORMATIONAL exchanges MUST ONLY occur after the initial exchanges and are cryptographically protected with the negotiated keys. |
| 34 | The IKE SA MUST be closed or rekeyed. |
| 35 | An endpoint MUST NOT conclude that the other endpoint has failed based on any routing information (e.g., ICMP messages) or IKE messages that arrive without cryptographic protection (e.g., Notify messages complaining about unknown SPIs), because these messages can be forged or sent by attackers. |

**Table 9: Properties extracted by LLM**